# Top-Ten Behavioral Biases in Project Management:
# An Overview

By Bent Flyvbjerg

First BT Professor and Inaugural Chair of Major Programme Management

University of Oxford's Saïd Business School

Villum Kann Rasmussen Professor and Chair of Major Program Management

IT University of Copenhagen

Professorial Fellow, St Anne's College, University of Oxford

## Abstract

Behavioral science has witnessed an explosion in the number of biases identified by behavioral scientists, to more than 200 at present. The paper identifies the ten most important behavioral biases for project planning and management. First, we argue it is a mistake to equate behavioral bias with cognitive bias, as is common. Cognitive bias is only half the story; political bias is the other half. Second, we list the top-ten behavioral biases in project management: strategic misrepresentation; optimism and uniqueness bias; the planning fallacy; overconfidence, availability, and hindsight bias; the base-rate fallacy; anchoring; and escalation of commitment. Each bias is defined and its impacts on project management explained, with examples. Third, base-rate neglect is identified as a primary reason projects underperform. This is supported by presentation of the most comprehensive set of base rates that exists in project management scholarship, from 2,062 projects. Finally, recent findings of power-law outcomes in project performance are identified as a possible first stage in discovering a general theory of project management, with more fundamental and more scientific explanations of project outcomes than found in existing theory.

*Keywords*: Behavioral science, behavioral economics, project planning, project management, behavioral bias, cognitive bias, political bias, strategic misrepresentation, optimism bias, uniqueness bias, planning fallacy, overconfidence bias, hindsight bias, availability bias, base-rate fallacy, anchoring, escalation of commitment, preferential attachment, power laws.

## Introduction

Since the early work of Amos Tversky and Daniel Kahneman, the number of biases identified by behavioral scientists has exploded in what has been termed a behavioral revolution in economics, management, and across the social and human sciences. Today, Wikipedia's list of cognitive biases contains more than two hundred items (Wikipedia 2021). The present paper gives an overview of the most important behavioral biases in project planning and management, summarized in Table 1. They are the biases most likely to trip up project planning and management and negatively impact project outcomes, if the biases are not identified and dealt with up front and during delivery.

Many would agree with Kahneman (2011: 255) that optimism bias "may well be the most significant of the cognitive biases." However, behavioral biases are not limited to cognitive biases, though behavioral scientists, and especially behavioral economists, often seem to think so. For instance, Nobel laureate Richard Thaler (2017: 261), godparent of behavioral economics, defines what he calls "the real point of behavioral economics" as "to highlight behaviors that are in conflict with the standard rational model." But nothing in this definition limits the object of behavioral economics to cognitive bias. Thaler (2017: 357) speaks of "the unrealism of hyperrational models," and we agree. But behavioral economics itself suffers from such unrealism, because it ignores that many behavioral phenomena are better explained by political bias than by cognitive bias.

Behavioral economics in its present form suffer from an overfocus on cognitive psychology: Economic decisions get overaccounted for in psychological terms, when other perspectives – for instance political, sociological, and organizational – may be more pertinent. If all you have is a hammer, everything looks like a nail. Similarly, if all you have is psychology, everything gets diagnosed as a psychological problem, even when it is not. Behavioral economics suffer from a "psychology bias," in this sense. Cognitive bias is therefore only half the story in behavioral science. Political bias is the other half.

Political bias – understood as deliberate strategic distortions – arises from power relations, instead of from cognition, and has long been the object of study in political economy. Political bias is particularly important for big, consequential decisions and projects, which are often subject to high political-organizational pressures. In fact, for very large projects – so-called megaprojects – the most significant behavioral bias is arguably political bias, more specifically strategic misrepresentation (Flyvbjerg et al. 2002, Wachs 2013, Flyvbjerg et al. 2018). Cognitive bias may account well for outcomes in the simple lab experiments done by behavioral scientists. But for real-world decision-making – in big hierarchical organizations, involving office politics, salesmanship, jockeying for position and funds, including in the C-suite and ministerial offices, with millions and sometimes billions of dollars at stake – political bias is pervasive and must be taken into account. Or so I argue.

*Table 1: Top-ten behavioral biases in project planning and management*

| Name of bias | Description |
|---|---|
| 1. Strategic misrepresentation | The tendency to deliberately and systematically distort or misstate information for strategic purposes. A.k.a. political bias, strategic bias, or power bias. |
| 2. Optimism bias | The tendency to be overly optimistic about the outcome of planned actions, including overestimation of the frequency and size of positive events and underestimation of the frequency and size of negative ones. |
| 3. Uniqueness bias | The tendency to see one's project as more singular than it actually is. |
| 4. Planning fallacy (writ large) | The tendency to underestimate costs, schedule, and risk and overestimate benefits and opportunities. |
| 5. Overconfidence bias | The tendency to have excessive confidence in one's own answers to questions. |
| 6. Hindsight bias | The tendency to see past events as being predictable at the time those events happened. Also known as the I-knew-it-all-along effect. |
| 7. Availability bias | The tendency to overestimate the likelihood of events with greater ease of retrieval (availability) in memory. |
| 8. Base-rate fallacy | The tendency to ignore generic base-rate information and focus on specific information pertaining to a certain case or small sample. |
| 9. Anchoring | The tendency to rely too heavily, or "anchor," on one trait or piece of information when making decisions, typically the first piece of information acquired on the relevant subject. |
| 10. Escalation of commitment | The tendency to justify increased investment in a decision, based on the cumulative prior investment, despite new evidence suggesting the decision may be wrong. Also known as the sunk-cost fallacy. |

It should be emphasized again that many other behavioral biases exist than those mentioned in Table 1, which are relevant to project planning and management, e.g., illusion of control, conservatism bias, normalcy bias, recency bias, probability neglect, the ostrich effect, and more. But the ten mentioned here may be considered the most important, in the sense they are deemed to be the most common biases with the most direct impact on project outcomes.

**Discussions with Kahneman**

My first opportunity to reflect systematically on the relationship between political and cognitive bias was an invitation in 2003 from the editor of *Harvard Business Review* (*HBR*) to comment on an article by Dan Lovallo and Daniel Kahneman (2003). The year before, Kahneman had won the Nobel Prize in Economics for his pathbreaking work with Amos Tversky (who died in 1996) on heuristics and biases in decision-making, including optimism bias, which was the topic of the *HBR* article. The editor explained to me that he saw Kahneman and me as explaining the same phenomena – cost overruns, delays, and benefit shortfalls in investment decisions – but with fundamentally different theories. As a psychologist, Kahneman explained outcomes in terms of cognitive bias, especially optimism bias and the planning fallacy. As an economic geographer, I explained the same phenomena in terms of political-economic bias, specifically strategic misrepresentation. So which of the two theories is right, asked the *HBR* editor?

The editor's question resulted in a spirited debate in the pages of *HBR*. I commented on Kahneman and Lovallo's article that they,

> "underrate one source of bias in forecasting – the deliberate 'cooking' of forecasts to get
> ventures started. My colleagues and I call this the Machiavelli factor. The authors
> [Kahneman and Lovallo] mention the organizational pressures forecasters face to
> exaggerate potential business results. But adjusting forecasts because of such pressures
> can hardly be called optimism or a fallacy; deliberate deception is a more accurate term.
> Consequently, Lovallo and Kahneman's analysis of the planning fallacy seems valid
> mainly when political pressures are insignificant. When organizational pressures are
> significant, both the causes and cures for rosy forecasts will be different from those
> described by the authors" (Flyvbjerg 2003:121)

Kahneman and Lovallo (2003: 122) responded:

> "Flyvbjerg and his colleagues reject optimism as a primary cause of cost overruns
> because of the consistency of the overruns over a significant time period. They assume
> that people, particularly experts, should learn not only from their mistakes but also from
> others' mistakes. This assumption can be challenged on a number of grounds."

Ultimately, the *HBR* debate did not so much resolve the question as clarify it and demonstrate its relevance. Kahneman and I therefore continued the discussion offline. Others have commented on Kahneman's generosity in academic interactions. He invited me to visit him at home, first in Paris and later New York, to develop the thinking on political and cognitive bias, and how they may be interrelated. He was more rigorous than anyone I'd discussed bias with before, and I found the discussions highly productive.

In addition to being generous, Kahneman is deeply curious and empirical. Based on our discussions, he decided he wanted to investigate political bias firsthand and asked if I could arrange for him to meet people exposed to such bias. I facilitated an interview with senior officials I knew at the Regional Plan Association of the New York-New Jersey-Connecticut metropolitan (tri-state) area, with offices near Kahneman's home in New York.[1] Their work includes forecasting and decision-making for major infrastructure investments in the tri-state region, which are among the largest and most complex in the world and which are the type of project I studied to develop my theories of strategic misrepresentation. Decision-making on such projects are a far cry from the lab experiments used by Kahneman and other behavioral scientists to document classic cognitive biases like loss aversion, anchoring, optimism, and the planning fallacy.

When Kahneman and I compared notes again, we agreed the balanced position regarding real-world decision-making is that both cognitive and political bias influence outcomes. Sometimes one dominates, sometimes the other, depending on what the stakes are and the degree of political-organizational pressures on individuals. If the stakes are low and political-organizational pressures absent, which is typical for lab experiments in behavioral science, then cognitive bias will dominate and such bias will be what you find. But if the stakes and pressures are high – for instance, when deciding whether to spend billions of dollars on a new subway line in Manhattan – political bias and strategic misrepresentation are likely to dominate and will be what you uncover, together with cognitive bias, which is hardwired and therefore present in most, if not all, situations.

Imagine a scale for measuring political-organizational pressures, from weak to strong. At the lower end of the scale, one would expect optimism bias to have more explanatory power of outcomes relative to strategic misrepresentation. But with more political-organizational pressures, outcomes would increasingly be explained in terms of strategic misrepresentation. Optimism bias would not be absent when political-organizational pressures increase, but optimism bias would be supplemented and reinforced by bias caused by strategic misrepresentation. Finally, at the upper end of the scale, with strong political-organizational pressures – e.g., the situation where a CEO or minister *must* have a certain project – one would expect strategic misrepresentation to have more explanatory power relative to optimism bias, again without optimism bias being absent. Big projects, whether in business or government, are typically at the upper end of the scale, with high political-organizational pressures and strategic misrepresentation. The typical project in the typical organization is somewhere in the middle of the scale, exposed to a mix of strategic misrepresentation and optimism bias, where it is not always clear which one is stronger.

The discussions with Kahneman taught me that although I had fully acknowledged the existence of cognitive bias in my original work on bias (Flyvbjerg et al. 2002) I needed to emphasize cognition more to get the balance right between political and psychological bias in real-life decision making. This was the object of later publications (Flyvbjerg et al. 2004, Flyvbjerg 2006, Flyvbjerg et al. 2009, Flyvbjerg 2013, Flyvbjerg et al. 2016). More importantly, however, in our discussions and in a relatively obscure paper by Kahneman and Tversky (1979a) I found an idea for how to eliminate or reduce both cognitive and political bias in decision-making. I developed this into a practical tool called "reference class forecasting" (Flyvbjerg 2006) and in *Thinking, Fast and Slow* Kahneman (2011: 251) was kind enough to endorse the method as an effective tool for bringing the outside view to bear on projects and decisions to de-bias them.

Finally, it has been encouraging to see Kahneman begin to mention political bias in his writings, including in his seminal book, *Thinking, Fast and Slow*, where he explicitly points out that,

> "Errors in the initial budget are not always innocent. The authors of unrealistic plans are
> often driven by the desire to get the plan approved – whether by their superiors or by a
> client – supported by the knowledge that projects are rarely abandoned unfinished
> merely because of overruns in costs or completion times" (Kahneman 2011: 250-51).

That is clearly not a description of cognitive bias, which is innocent per definition, but of political bias, specifically strategic misrepresentation aimed at getting projects underway.

Most likely, none of the above would have happened without the *HBR* editor's simple question, "Strategic misrepresentation or optimism bias, which is it?" The discussions with Kahneman proved the answer to be: "Both."

Below we use this insight to describe the most important behavioral biases in project planning and management, starting with strategic misrepresentation, followed by optimism bias and eight other biases.

## 1. Strategic Misrepresentation

Strategic misrepresentation is the tendency to deliberately and systematically distort or misstate information for strategic purposes (Jones and Euske 1991, Steinel and De Dreu 2004). This bias is sometimes also called political bias, strategic bias, power bias, or the Machiavelli factor (Guinote and Vescio 2010). The bias is a rationalization where the ends justify the means. The strategy (e.g., achieve funding) dictates the bias (e.g., make projects look good on paper). Strategic misrepresentation can be traced to agency problems and political-organizational pressures, for instance competition for scarce funds or jockeying for position. Strategic misrepresentation is deliberate deception, and as such it is lying, per definition (Bok 1999, Carson 2006, Fallis 2009).

Here, a senior Big-Four consultant explains how strategic misrepresentation works in practice:

> "In the early days of building my transport economics and policy group at [name of
> company omitted], I carried out a lot of feasibility studies in a subcontractor role to
> engineers. In virtually all cases it was clear that *the engineers simply wanted to justify the project
> and were looking to the traffic forecasts to help in the process* ... I once asked an engineer why their cost estimates were invariably underestimated and he simply answered '*if we gave the true expected outcome costs nothing would be built*'" (personal communication, author's archives, italics added).

Signature architecture is notorious for large cost overruns. A leading signature architect, France's Jean Nouvel, winner of the Pritzker-Prize, explains how it works:

"I don't know of buildings that cost less when they were completed than they did at the outset. In France, there is often a theoretical budget that is given because it is the sum that politically has been released to do something. In three out of four cases this sum does not correspond to anything in technical terms. *This is a budget that was made because it could be accepted politically. The real price comes later*. The politicians make the real price public where they want and when they want" (Nouvel 2009: 4, italics added).

This is strategic misrepresentation. Following its playbook, a strategic cost or schedule estimate will be low, because it is more easily accepted, leading to cost and schedule overrun. Similarly, a strategic benefit estimate will be high, leading to benefit shortfalls. Strategic misrepresentation therefore produces a systematic bias in outcomes. And this is precisely what the data show (see Table 2).[2] We see the theory of strategic misrepresentation fits the data. Explanations of project outcomes in terms of strategic misrepresentation have been set forth by Wachs (1989, 1990, 2013), Kain (1990), Pickrell (1992), and Flyvbjerg et al. (2002, 2004, 2005, 2009), among others.

Strategic misrepresentation will be particularly strong where political-organizational pressures are high, as argued above, and such pressures are especially high for big, strategic projects. The bigger and more expensive the project, the more strategic import it is likely to have with more attention from top management and with more opportunities for political-organizational pressures to develop, other things being equal. For project planning and management the following propositions apply:

*Proposition 1*: For small projects, with low strategic import and no attention from top management, bias, if present, is likely to originate mainly with cognitive bias, e.g., optimism bias.

*Proposition 2:* For big projects, with high strategic import and ample attention from top management, bias, if present, is likely to originate mainly with political bias, e.g., strategic misrepresentation, although cognitive bias is also likely to be present.

Strategic misrepresentation has proved especially important in explaining megaproject outcomes. For megaproject management, strategic misrepresentation may be expected to be the dominant bias (Flyvbjerg 2014).

Professor Martin Wachs of UC Berkeley and UCLA, who pioneered research on strategic misrepresentation in transportation infrastructure forecasting, recently looked back at more than 25 years of scholarship in the area. After carefully weighing the evidence for and against different types of explanations of forecasting inaccuracy, Wachs summarized his findings in the following manner:

> "While some scholars believe this [misleading forecasting] is a simple technical matter involving the tools and techniques of cost estimation and patronage forecasting, there is growing evidence that the gaps between forecasts and outcomes are the results of deliberate misrepresentation and thus amount to a collective failure of professional ethics ... Often ... firms making the forecasts stand to benefit if a decision is made to proceed with the project" (Wachs 2013: 112).

Wachs found a general incentive to misrepresent forecasts for infrastructure projects and that this incentive drives forecasting outcomes. Wachs's review and the studies cited above falsify the notion that optimism and other cognitive biases may serve as a stand-alone explanation of cost underestimation and benefit overestimation, which has been the common view in behavioral science. Explanations in terms of cognitive bias are especially wanting in situations with high political and organizational pressures. In such situations forecasters, planners, and decision makers intentionally use the following Machiavellian formula to make their projects look good on paper, with a view to securing their approval and funding:

Underestimated costs + Overestimated benefits = Funding

Finally, recent research has found that not only do political and cognitive biases compound each other in the manner described above. Experimental psychologists have shown that political bias directly amplifies cognitive bias in the sense that people who are powerful are affected more strongly by various cognitive biases – e.g., availability bias and recency bias – than people who are not (Weick and Guinote 2008). A heightened sense of power also increases individuals' optimism in viewing risks and their propensity to engage in risky behavior (Anderson and Galinsky 2006: 529). This is because people in power tend to disregard the rigors of deliberate rationality, which are too slow and cumbersome for their purposes. They prefer – consciously or not – subjective experience and intuitive judgment as the basis for their decisions, as documented by Flyvbjerg (1998: 69 ff.), who found that people in power will deliberately exclude experts from meetings when much is at stake, in order to avoid clashes in high-level negotiations between power's intuitive decisions and experts' deliberative rationality. Guinote and Vescio (2010) similarly found that people in power rely on ease of retrieval more than people without power. In consequence, total bias -- political plus cognitive -- escalates, but not in a simple linear manner where total bias equals the sum of political and cognitive biases, but instead in a complex, convex way where political bias amplifies cognitive bias, leading to convex risk. This, undoubtedly, is one reason we find strong convexities in the planning and management of big projects. Decisions about big projects are typically made by highly powerful people, and such individuals are convexity generators, with political bias driving their cognitive biases, which are larger for powerful individuals than for non-powerful ones.

## 2. Optimism Bias

Optimism bias is a cognitive bias and it is the tendency for individuals to be overly bullish about the outcomes of planned actions (Kahneman 2011: 255). Sharot (2011: xv) calls it "one of the greatest deceptions of which the human mind is capable." Where strategic misrepresentation is deliberate, optimism bias is non-deliberate. In the grip of optimism, people – including experts – are unaware they are optimistic. They make decisions based on an ideal vision of the future rather than on a rational weighting of gains, losses, and probabilities. They overestimate benefits and underestimate costs. They involuntarily spin scenarios of success and overlook the potential for mistakes and miscalculations. As a result, plans are unlikely to deliver as expected in terms of benefits and costs.

Almost a hundred years ago, when Geoffrey Faber founded what would become Faber & Faber, the renowned London publishing house, he was so certain of his project that he bet his mother's, his own, and a few friends' fortune on it, concluding, "everybody would benefit ... with a substantial income" (Faber 2019: 6, underline in the original). A year later, the new publishing house was in its first of several near-bankruptcies, and Faber wrote in his diary:

> "I find it hard to justify my buoyant self-confidence of last year ... I ought, I think, to
> have foreseen trouble and gone more cautiously" (Faber 2019: 27-28).

That's optimism bias and what it does to individuals. Geoffrey Faber is not the only entrepreneur to have been tripped up like this. It's typical. What's less typical is that Faber & Faber survived to tell the story. Most companies fail and are forgotten.

Optimism bias can be traced to cognitive biases, i.e., systematic deviations from rationality in the way the mind processes information (Shepperd et al. 2002, Sharot et al. 2007, O'Sullivan 2015). These biases are thought to be ubiquitous. In project planning and management, an optimistic cost or schedule estimate will be low, leading to cost and schedule overrun. An optimistic benefit estimate will be high, leading to benefit shortfalls. Optimism therefore produces a systematic bias in project outcomes, which is what the data show (see Table 2). The theory of optimism bias thus fit the data, which lends support to its validity.

Interestingly, however, when researchers ask forecasters about causes of inaccuracies in their forecasts, they do not mention optimism bias as a main cause, whereas they do mention strategic misrepresentation and the usual suspects: scope changes, complexity, price changes, unexpected underground conditions, bad weather, etc. (Flyvbjerg et al. 2005: 138-140). Psychologists would argue this is because optimism bias is a true cognitive bias. As such it is unreflected by forecasters, including when they participate in surveys about causes of forecasting inaccuracy. Psychologists would further argue there is a large body of experimental evidence for the existence of optimism bias (Buehler et al., 1994, 1997; Newby-Clark et al. 2000). However, the experimental data are mostly from simple laboratory experiments with students. This is a problem, because it's an open question to what extent the results apply outside the laboratory, in real-life situations like project planning and management.

Optimism bias can be both a blessing and a curse. Optimism and a "can-do" attitude are obviously necessary to get projects done. Kahneman (2011: 255) calls optimism "the engine of capitalism." I would go further and call it the engine of life. But optimism can seriously trip us up if we are unaware of its pitfalls and therefore take on risks we would have avoided had we known the real, non-optimistic, odds. This has been known and reflected since at least the ancient Greeks. More than two millennia ago, the Greek historian Thucydides (2009: 220) said about the Athenians that "they expected no reverses" to "their current good fortune" -- i.e., they were optimistic, specifically overconfident -- and this caused the fall of Athens in the Peloponnesian War, according to Thucydides.

No upside can compensate for the ultimate downside, death. This is a fundamental asymmetry between upside and downside in human existence and is probably why humans are predisposed for loss aversion, as documented by prospect theory (Kahneman and Tversky 1979b). Quite simply, it is rational in evolutionary terms to be more concerned about downside than upside. "Death" does not have to be of an individual, needless to say. It can be of a nation, a city, a business, or a project.

In my research I have found that successful leaders have a rare combination of hyper realism and can-do optimism (Flyvbjerg and Gardner, forthcoming). I call such individuals "realistic optimists." Risto Siilasmaa, chairman of Nokia during its recent successful turnaround, goes one step further in highlighting the two disparate dispositions, when he emphasizes "paranoid optimism" as the key to success in leading projects and businesses, always planning for the worst-case scenario: "The more paranoid we are, the harder we will continue to labor to shift the probability curve in our favor and the more optimistic we can afford to be" (Siilasmaa 2018: xvi). If you are looking for someone to successfully lead a project, this is the type of person you want: a realistic optimist, if not a paranoid one. You would never get on a plane if you overheard the pilot say to the co-pilot, "I'm optimistic about the fuel-situation." Similarly, one should not trust a project leader who is optimistic about the budget or schedule, which is the fuel of projects.

During the Apollo program (1961–1972), the NASA administration criticized its cost engineers for being optimistic with a US$10 billion estimate for the program (approximately US$90 billion in 2021 dollars). The administration told the engineers that their assumption "that everything's going to work" was wrong (Bizony, 2006: 41). The engineers then increased their estimate to US$13 billion, which the administration adjusted to US$20 billion and got approved by Congress, to the shock of the engineers. Today, the NASA administration's US$7 billion increase has a technical name: "optimism bias uplift." NASA jokingly called it the "administrator's discount." But they were serious when they advised that all senior executives in charge of large, complex projects must apply such a discount to make allowance for the unknown. Whatever the name, it is the single most important reason Apollo has gone down in history as that rare species of multi-billion-dollar project: one delivered on budget. The NASA administration "knew exactly what [it] was doing" with Apollo, as rightly observed by space historian Piers Bizony (ibid.).

Explanations of project outcomes in terms of optimism bias originate with Kahneman and Tversky (1979a) and have been further developed by Kahneman and Lovallo (1993), Lovallo and Kahneman (2003), Flyvbjerg (2009a), and Flyvbjerg et al. (2009).

Above we saw that strategic project planners and managers sometimes underestimate cost and overestimate benefit to achieve approval for their projects. Optimistic planners and managers also do this, albeit non-intentionally. The result is the same, however, namely cost overruns and benefit shortfalls. Thus optimism bias and strategic misrepresentation reinforce each other, when both are present in a project. An interviewee in our research described this strategy as "showing the project at its best" (Flyvbjerg et al. 2004: 50). It results in an inverted Darwinism, i.e., "survival of the unfittest" (Flyvbjerg 2009b). It is not the best projects that get implemented like this, but the projects that look best on paper. And the projects that look best on paper are the projects with the largest cost underestimates and benefit overestimates, other things being equal. But the larger the cost underestimate on paper, the greater the cost overrun in reality. And the larger the overestimate of benefits, the greater the benefit shortfall. Therefore, the projects that have been made to look best on paper become the worst, or unfittest, projects in reality.

## 3. Uniqueness Bias

Uniqueness bias was originally identified by psychologists as the tendency of individuals to see themselves as more singular than they actually are, e.g., singularly healthy, clever, or attractive (Suls and Wan 1987, Suls et al. 1988, Goethals et al. 1991). In project planning and management, the term was first used by Flyvbjerg (2014: 9), who defined uniqueness bias as the tendency of planners and managers to see their projects as singular. It is a general bias, but it turns out to be particularly rewarding as an object of study in project management, because project planners and managers are systematically primed to see their projects as unique.

The standard definition of a project, according to the biggest professional organization in the field, the US-based Project Management Institute (PMI 2017: 4), directly emphasizes uniqueness as one of two defining features of what a project is: "A project is a temporary endeavor undertaken to create a *unique* product, service, or result" (italics added). Similarly, the UK-based Association for Project Management (APM 2012) stresses uniqueness as the very first characteristic of what a project is in their official definition: "A project is a *unique*, transient endeavour, undertaken to achieve planned objectives " (italics added). Academics, too, define projects in terms of uniqueness, here Turner and Müller (2003: 7, italics added): "A project is a temporary organization to which resources are assigned to undertake a *unique*, novel and transient endeavour managing the inherent uncertainty and need for integration in order to deliver beneficial objectives of change" (Turner and Müller 2003: 7, italics added). Similar views of uniqueness as key to the nature of projects may be found in Grün (2004: 3, 245), Fox and Miller (2006: 3, 109), and Merrow (2011: 161).

We maintain that the understanding of projects as unique is unfortunate, because it contributes to uniqueness bias with project planners and managers. In the grip of uniqueness bias, project managers see their projects as more singular than they actually are. This is reinforced by the fact that new projects often use non-standard technologies and designs.

Uniqueness bias tends to impede managers' learning, because they think they have little to learn from other projects as their own project is unique. Uniqueness bias may also feed overconfidence bias (see below) and optimism bias (see above), because planners subject to uniqueness bias tend to underestimate risks. This interpretation is supported by research on IT project management reported in Flyvbjerg and Budzier (2011), Budzier and Flyvbjerg (2013), and Budzier (2014). The research found that managers who see their projects as unique perform significantly worse than other managers. If you are a project leader and you overhear team members speak of your project as unique, you therefore need to react.

It is self-evidently true, of course, that a project may be unique in its own specific geography and time. For instance, California has never built a high-speed rail line before, so in this sense the California High-Speed Rail Authority is managing a unique project. But the project is only unique to California, and therefore not truly unique. Dozens of similar projects have been built around the world, with data and lessons learned that would be highly valuable to California. In that sense projects are no different from people. A quote, often ascribed to the anthropologist Margaret Mead, captures the point well: "Always remember that you are absolutely unique. Just like everyone else." Each person is obviously unique, but also has a lot in common with other people. The uniqueness of people has not stopped the medical profession from making progress based on what humans have in common. The problem with project management is that uniqueness bias hinders such learning across projects, because project managers and scholars are prone to "localism bias," which we define as the tendency to see the local as global, due to availability bias for the local. Localism bias explains why local uniqueness is easily and often confused with global uniqueness. In many projects it does not even occur to project planners and managers to look outside their local project, because "our project is unique," which is a mantra one hears over and over in projects, and which it is surprisingly easy to get project managers to admit to.

Uniqueness bias feeds what Kahneman (2011: 247) calls the "inside view." Seeing things from this perspective, planners focus on the specific circumstances of the project they are planning and seek evidence in their own experience. Estimates of budget, schedule, etc. are based on this information, typically built "from the inside and out," or bottom-up, as in conventional cost engineering. The alternative is the "outside view," which consists in viewing the project you are planning from the perspective of similar projects that have already been completed, basing your estimates for the planned project on the actual outcomes of these projects. But if your project is truly unique then similar projects clearly do not exist, and the outside view becomes irrelevant and impossible. This leaves you with the inside view as the only option for planning your project. Even if a project is not truly unique, if the project team thinks it is then the outside view will be left by the wayside and the inside view will reign supreme, which is typical. "In the competition with the inside view, the outside view does not stand a chance," as pithily observed by Kahneman (2011: 249). The inside view is the perspective people spontaneously adopt when they plan, reinforced by uniqueness bias for project planners and managers. The inside view is therefore typical of project planning and management. The consequences are dire, because only the outside view effectively takes into account all risks, including the so-called "unknown unknowns." These are impossible to predict from the inside, because there are too many ways a project can go wrong. However, the unknown unknowns are included in the outside view, because anything that went wrong with the completed projects that constitute the outside view is included in their outcome data (Flyvbjerg 2006). Using these data for planning and managing a new project therefore leaves you with a measure of all risk, including unknown unknowns. Uniqueness bias makes you blind to unknown unknowns. The outside view is an antidote to uniqueness bias.

For project managers, in addition to being predisposed, like everyone else, for the inside view and uniqueness, they have been indoctrinated by their professional organizations to believe projects are unique, as we saw above. Thus it's no surprise it takes substantial experience to cut loose from the conventional view. Patrick O'Connell, an experienced megaproject manager and Practitioner Director of Oxford's BT Centre for Major Programme Management, told me, "The first 20 years as a megaproject manager I saw uniqueness in each project; the next 20 years similarities." The NASA administration, mentioned above, balked when people insisted the Apollo program, with its aim of landing the first humans on the moon, was unique. How could it not be, as putting people on the moon had never been done before, people argued. The administration would have none of it. They deplored those who saw the program "as so special—as so exceptional," because such people did not understand the reality of the project. The administration insisted, in contrast, that "the basic knowledge and technology and the human and material resources necessary for the job already existed," so there was no reason to reinvent the wheel (Webb, 1969, p. 11, p. 61). The NASA-Apollo view of uniqueness bias saw this bias for what it is: a fallacy.

In sum, uniqueness bias feeds the inside view and optimism, which feeds underestimation of risk, which make project teams take on risks they would likely not have accepted had they known the real odds. Good project leaders do not let themselves be fooled like this. They know PMI and APM are wrong when they say projects are unique. Projects are often unique locally, yes. But to be locally unique is an oxymoron. This, however, is typically the meaning of the term "unique," when used in project management. It is a misnomer that undermines project performance and thus the project management profession. Truly unique projects are rare. We have lots to learn from other projects, always. And if we don't learn, we will not succeed with our projects.

## 4. The Planning Fallacy (Writ Large)

The planning fallacy is a subcategory of optimism bias that arises from individuals producing plans and estimates that are unrealistically close to best-case scenarios. The term was originally coined by Kahneman and Tversky (1979a: 315) to describe the tendency for people to underestimate task-completion times. Buehler et al. (1994, 1997) continued work following this definition. Later, the concept was broadened to cover the tendency for people to, on the one hand, underestimate costs, schedules, and risks for planned actions and, on the other, overestimate benefits and opportunities for those actions. Because the original narrow and later broader concepts are so fundamentally different in the scope they cover, Flyvbjerg and Sunstein (2017) suggested the term "planning fallacy writ large" for the broader concept, to avoid confusing the two.

Flyvbjerg et al. (2003: 80) call the tendency to plan according to best-case scenarios the "EGAP-principle," for Everything Goes According to Plan. The planning fallacy and the EGAP-principle are similar in the sense that both result in a lack of realism, because of their overreliance on best-case scenarios, as with the NASA cost engineers above. Both lead to base-rate neglect, illusion of control, and overconfidence. In this manner, both feed into optimism bias.

At the most fundamental level, Kahneman and Tversky identified the planning fallacy as arising from a tendency with people to neglect distributional information when they plan. People who plan would adopt what Kahneman and Tversky (1979a: 315) first called an "internal approach to prediction," and later renamed the "inside view," under the influence of which people would focus on "the constituents of the specific problem rather than on the distribution of outcomes in similar cases." Kahneman and Tversky (ibid.) emphasized that, "The internal approach to the evaluation of plans is likely to produce underestimation [of schedules]." For the planning fallacy writ large, such underestimation applies to costs, schedules, and risk, whereas overestimation applies to benefits and opportunities.

Interestingly, Guinote (2017: 365-66) found in an experiment that subjects who had been made to feel in power were more likely to underestimate the time needed to complete a task than those not in power, demonstrating a higher degree of planning fallacy for people in power. Again, this is an example of how power bias and cognitive bias interact, resulting in amplification and convexity.

The planning fallacy's combination of underestimated costs and overestimated benefits generates risks to the second degree. Instead of cost risk and benefit risk cancelling out one another – as other theories predict, e.g., Hirschman's (2014) principle of the Hiding Hand – under the planning fallacy the two types of risk reinforce each other, creating convex (accelerated) risks for projects from the get-go. The planning fallacy goes a long way in explaining the Iron Law of project management: "Over budget, over time, under benefits, over and over again" (Flyvbjerg 2017). As a project leader, you want to avoid convex risks because such risks are particularly damaging. You want to avoid committing the planning fallacy, and especially for people in power.

## 5-7. Overconfidence Bias, Hindsight Bias, and Availability Bias

Overconfidence bias is the tendency to have excessive confidence in one's own answers to questions, and to not fully recognize the uncertainty of the world and one's ignorance of it. People have been shown to be prone to what is called the "illusion of certainty" in (a) overestimating how much they understand and (b) underestimating the role of chance events and lack of knowledge, in effect underestimating the variability of events they are exposed to in their lives (Pallier et al. 2002, Moore and Healy 2008, Proeger and Meub 2014). Overconfidence bias is found with both laypeople and experts, including project planners and managers (Fabricius and Büttgen 2015).

Overconfidence bias is fed by illusions of certainty, which are fed by hindsight bias, also known as the "I-knew-it-all-along effect." Availability bias – the tendency to overweigh whatever comes to mind – similarly feeds overconfidence bias. Availability is influenced by the recency of memories and by how unusual or emotionally charged they may be, with more recent, more unusual, and more emotional memories being more easily recalled. Overconfidence bias is a type of optimism and it feeds overall optimism bias.

A simple way to illustrate overconfidence bias is to ask people to estimate confidence intervals for statistical outcomes. In one experiment, the Chief Financial Officers (CFOs) of large US corporations were asked to estimate the return next year on shares in the relevant Standard & Poor's index (Kahneman 2011: 261). In addition, the CFOs were asked to give their best guess of the 80 percent confidence interval for the estimated returns by estimating a value for returns they were 90 percent sure would be too low (the lower decile, or P10) and a second value they were 90 percent sure would be too high (the upper decile, or P90), with 80 percent of returns estimated to fall between these two values (and 20 percent outside). Comparing actual returns with the estimated confidence interval, it was found that 67 percent of actual returns fell outside the estimated 80-percent confidence interval, or 3.35 times as many as estimated. The actual variance of outcomes was grossly underestimated by these financial experts, which is the same as saying they grossly underestimated risk. It is a typical finding. The human brain, including the brains of experts, spontaneously underestimates variance. For whatever reason, humans seem hardwired for this.

In project management, overconfidence bias is unfortunately built into the very tools experts use for quantitative risk management. The tools, which are typically based on computer models using so-called Monte-Carlo simulations, or similar, look scientific and objective, but are anything but. Again, this is easy to document. You simply compare assumed variance in a specific, planned project with actual, historic variance for its project type, and you find the same result as for the CFOs above (Batselier and Vanhoucke 2016). The bias is generated by experts assuming thin-tailed distributions of risk (normal or near-normal), when the real distributions are fat-tailed (lognormal, power law, or similar probability distribution) (Taleb 2004). The error is not with Monte-Carlo models as such, but with erroneous input into the models. Garbage in, garbage out, as always. To eliminate overconfidence bias you want a more objective method that takes all distributional information into account, not just the distributional information experts can think of, which is subject to availability bias. The method needs to run on historical data from projects that have actually been completed. Flyvbjerg (2006) describes such a method.

In the thrall of overconfidence bias, project planners and decision makers underestimate risk by overrating their level of knowledge and ignoring or underrating the role of chance events in deciding the fate of projects. Hiring experts will generally not help, because experts are just as susceptible to overconfidence bias as laypeople and therefore tend to underestimate risk, too. There is even evidence that the experts who are most in demand are the most overconfident. I.e., people are attracted to, and willing to pay for, confidence, more than expertise (Kahneman 2011: 263, Tetlock 2005). Risk underestimation feeds the Iron Law of project management and is the most common cause of project downfall. Good project leaders must know how to avoid this.

Individuals produce confidence by storytelling. The more coherent a story we can tell about what we see, the more confident we feel. But coherence does not necessarily equal validity. People tend to assume "what you see is all there is," called WYSIATI by Kahneman (2011: 87-88), who gives this concept pride of place in explaining a long list of biases, including overconfidence bias. People spin a story based on what they see. Under the influence of WYSIATI they spontaneously impose a coherent pattern on reality, while they suppress doubt and ambiguity and fail to allow for missing evidence, according to Kahneman. The human brain excels at inferring patterns and generating meaning based on skimpy, or even non-existent, evidence. But coherence based on faulty or insufficient data is not true coherence, needless to say. If we are not careful, our brains quickly settle for anything that looks like coherence and uses it as a proxy for validity. This may not be a big problem most of the time, and may even be effective, on average, in evolutionary terms, which may be why the brain works like this. But for big consequential decisions, typical of project planning and management, it is not an advisable strategy. Nevertheless, project leaders and their teams often have a very coherent – and very wrong – story about their project, for instance that the project is unique, as we saw above under uniqueness bias, or that the project may be completed faster and cheaper than the average project, or that everything will go according to plan. The antidote is better, more carefully curated stories, based on better data.

Gigerenzer (2018: 324) has rightly observed that overconfidence, presented by psychologists as a non-deliberate cognitive bias, is in fact often a deliberate strategic bias used to achieve predefined objectives, i.e., it is strategic misrepresentation. Financial analysts, for instance, "who earn their money by mostly incorrect predictions such as forecasting exchange rates or the stock market had better be overconfident; otherwise few would buy their advice," argues Gigerenzer, who further observes about this fundamental confusion of one type of bias for a completely different one that, "[c]onceptual clarity is desperately needed" (ibid.).

Finally, regarding the relationship between power bias and cognitive bias mentioned above, powerful individuals have been shown to be more susceptible to availability bias than individuals who are not powerful. The causal mechanism seems to be that powerful individuals are affected more strongly by ease of retrieval than by the content they retrieve, because they are more likely to "go with the flow" and trust their intuition than individuals who are not powerful (Weick and Guinote 2008). This finding has been largely ignored by behavioral economists, including by Thaler (2015) in his history of the field. This is unfortunate, because the finding documents convexity to the second degree for situations with power. By overlooking this, behavioral economists make the same mistake they criticize conventional economists for, namely overlooking and underestimating variance and risk. Conventional economists make the mistake by disregarding cognitive bias; behavioral economists by ignoring power bias and its effect on cognitive bias. Underestimating convexity is a very human mistake, to be sure. We all do it. But it needs to be accounted for if we want to understand all relevant risks and protect ourselves against them in project planning and management.

## 8. The Base-Rate Fallacy

The base-rate fallacy – sometimes also called base-rate bias or base-rate neglect – is the tendency to ignore base-rate information (general data pertaining to a statistical population or a large sample, e.g., its average) and focus on specific information (data only pertaining to a certain case, or a small number of cases) (Bar-Hillel 1980, Tversky and Kahneman 1982). If you play poker and assume different odds than those that apply, you are subject to the base-rate fallacy, and likely to lose. The objective odds are the base rate.

People often think the information they have is more relevant than it actually is, or they are blind to all the relevant information they do not have. Both situations result in the base-rate fallacy. "Probability neglect" – a term coined by Sunstein (2002: 62-63) to denote the situation where people overfocus on bad outcomes with small likelihoods, for instance terrorist attacks – is a special case of the base-rate fallacy.

The base-rate fallacy is fed by other biases, for instance uniqueness bias, described above, which results in extreme base-rate neglect, because the case at hand is believed to be unique, wherefore information about other cases is deemed irrelevant. The inside view, hindsight bias, availability bias, recency bias, WYSIATI bias, overconfidence bias, and framing bias also feed the base-rate fallacy. Base-rate neglect is particularly pronounced when there is a good, strong story. Big, monumental projects typically have such a story, contributing to extra base-rate neglect for those. Finally, we saw above that people, including experts, underestimate variance. In the typical project, base-rate neglect therefore combines with variation neglect, along the following formula:

Base-rate neglect + variation neglect = strong convexity

Preliminary results from our research indicate that variation neglect receives less attention in project management than base-rate neglect, which is unfortunate, because the research also indicates that variation neglect is typically larger and has even more drastic impact on project outcomes than base-rate neglect.

The base-rate fallacy runs rampant in project planning and management, as documented by the Iron Law described earlier. Table 2 shows the most comprehensive overview that exists of base rates for costs and benefits in project management, based on data from 2,062 projects covering eight project types. Most projects do not get base rates right – not even close, as documented by averages that are different from one (1.0 ≈ correct base rate) at a level of statistical significance so high (p <

0.0001 ≈ "overwhelmingly high," in statisticians' language) it is rarely found in studies of human behavior. The base-rate fallacy is deeply entrenched in project management, the data show. Flyvbjerg and Bester (forthcoming) argue that base-rate neglect results in a new behavioral bias, which they call the "cost-benefit fallacy," which routinely derail cost-benefit analyses of projects to a degree where such analyses cannot be trusted.

*Table 2: Base rates for cost and benefit overrun in 2,062 capital investment projects across eight types. Project planners and managers clearly do not get base rates right. The data show strong biases for (a) cost underestimation and overrun and (b) benefit overestimation and shortfall. Overrun is measured as actual divided by estimated costs and benefits (A/E), respectively, in real terms, baselined at the final investment decision.*

| Investment type | Cost overrun (A/E) | | | Benefit overrun (A/E) | | |
|---|---|---|---|---|---|---|
| | **N** | **Average** | **p\*** | **N** | **Average** | **p\*** |
| Dams | 243 | 1.96 | < 0.0001 | 84 | 0.89 | < 0.0001 |
| BRT† | 6 | 1.41 | 0.031 | 4 | 0.42 | 0.12 |
| Rail | 264 | 1.40 | < 0.0001 | 74 | 0.66 | < 0.0001 |
| Tunnels | 48 | 1.36 | < 0.0001 | 23 | 0.81 | 0.03 |
| Power plants | 100 | 1.36 | 0.0076 | 23 | 0.94 | 0.11 |
| Buildings | 24 | 1.36 | 0.00087 | 20 | 0.99 | 0.77 |
| Bridges | 49 | 1.32 | 0.00012 | 26 | 0.96 | 0.099 |
| Roads | 869 | 1.24 | < 0.0001 | 532 | 0.96 | < 0.0001 |
| **Total** | **1603** | **1.39/1.43††** | **< 0.0001** | **786** | **0.94/0.83††** | **< 0.0001** |

\*) The p-value of Wilcoxon test with null hypothesis that the distribution is symmetrically centered around one.
†) Bus rapid transit.
††) Weighted and unweighted average, respectively.
Note: See Flyvbjerg (2016: 181-182) for a description of the dataset used in the table.

As pointed out by Kahneman (2011: 150), "anyone who ignores base rates and the quality of evidence in probability assessments will certainly make mistakes." The cure to the base-rate fallacy, in and out of project management, is to get the base rate right by taking an outside view, for instance through reference class forecasting, carrying out premortems, or doing decision hygiene (Flyvbjerg 2006, Klein 2007, Kahneman et al. 2011, 2021: 312-324, 371-72).

If you're a project planner or manager, the easiest and most effective way to get started with curbing behavioral biases in your work is getting your base rates right, for the projects you're working on. Hopefully, most can see that if you don't understand the real odds of a game, you're unlikely to succeed at it. But that's the situation for most project planners and managers: they don't get the odds right for the game they are playing: project management. Table 2 documents this beyond reasonable doubt, and establishes realistic base rates for a number of important areas in project management that planners can use as a starting point for getting their projects right. Data for other project types were not included for reasons of space but show similar results.

### 9. Anchoring

Anchoring is the tendency to rely too heavily, or "anchor," on one piece of information when making decisions. Anchoring was originally demonstrated and theorized by Tversky and Kahneman (1974). In their perhaps most famous experiment, subjects were asked to estimate the percentage of African countries in the United Nations. First, a number between 0 and 100 was determined by spinning a wheel of fortune in the subjects' presence. Second, the subjects were instructed to indicate whether that number was higher or lower than the percentage of African countries in the United Nations. Third, the subjects were asked to estimate this percentage by moving upward or downward from the given number. The median estimate was 25 percent for subjects who received the number 10 from the wheel of fortune as their starting point, whereas it was 45 percent for subjects who started with 65. A random anchor significantly influenced the outcome.

Similar results have been found in other experiments for a wide variety of different subjects of estimation (Chapman and Johnson 1999, Fudenberg et al. 2012). Anchoring is pervasive. The human brain will anchor in most anything, whether random numbers, previous experience, or false information. It has proven difficult to avoid this (Wilson et al. 1996, Epley and Gilovich 2006, Simmons et al. 2010). The most effective way of dealing with anchoring therefore seems to be, not to avoid it, but to make sure the brain anchors in relevant information before making decisions, e.g., in base rates that are pertinent to the decision at hand, as proposed by Flyvbjerg (2006). This advice is similar to recommending gamblers must know the objective odds of each game they play, to increase their chances of winning and limit their losses. It is sound advice, but often goes unheeded.

Project planners and managers tend to err by anchoring their decisions in plans that are best-case, instead of most-likely, scenarios, as mentioned above. Planners and organizations also frequently anchor in their own limited experience, instead of seeking out a broader scope of histories, which would be more representative of the wider range of possible outcomes that actually apply to the project they are planning.

This happened to Hong Kong's MTR Corporation when they were tasked with building the first high-speed rail line in the territory. MTR anchored in its own experience with urban and conventional rail instead of throwing the net wider and looking at high-speed rail around the world. High-speed rail is significantly more difficult to build than urban and conventional rail and MTR had never built a high-speed rail line before. Despite – or perhaps because of – MTR's proven competence in building urban and conventional rail, the anchor for the high-speed rail line proved optimistic, resulting in significant cost and schedule overruns for the new venture (Flyvbjerg et al. 2014).

Ansar et al. (2014: 48) similarly found that planners of large dams around the world have generally anchored in the North-American experience with building dams, for no better reason than North America built their dams first. By choosing this anchor, planners ended up allowing insufficient adjustments to fully reflect local risks, e.g., exchange rate risks, corruption, logistics, and the quality of local project management teams. This resulted in optimistic budgets and higher cost overruns for dams built outside North America.

Anchoring is fed by other biases, including availability bias and recency bias, which induce people to anchor in the most available or most recent information, respectively. Anchoring results in base-rate neglect, i.e., underestimation of the probabilities, and thus the risks, that face a project (see previous section). Smart project leaders avoid this by anchoring their project in the base rate for similar projects to the one they are planning, for instance by benchmarking their project against outcomes for a representative class of similar, completed projects. Kahneman (2011: 154) explicitly identifies anchoring in the base rate as the cure for the WYSIATI bias mentioned above. Anchoring in the base rate is similar to taking an outside view, and the outside view is "an anchor that is meaningful," as rightly observed by Tetlock and Gardner (2015: 117-120), whereas spontaneous anchors typically are less meaningful and lead to biased decisions with hidden risks.

## 10. Escalation of Commitment

Last, but not least, escalation of commitment (sometimes also called commitment bias) is the tendency to justify increased investment in a decision, based on the cumulative prior investment, despite new evidence suggesting the decision may be wrong and additional costs will not be offset by benefits. Consider the example of two friends with tickets for a professional basketball game a long drive from where they live. On the day of the game there is a big snowstorm. The higher the price the friends paid for the tickets, the more likely they are to brave the blizzard and attempt driving to the game, investing more time, money, and risk (Thaler 2015: 20). That's escalation of commitment. In contrast, the rational approach when deciding whether to invest further in a venture would be to disregard what you have already invested.

Escalation of commitment applies to individuals, groups, and whole organizations. It was first described by Staw (1976) with later work by Brockner (1992), Staw (1997), Sleesman et al. (2012), and Drummond (2014, 2017). Economists use related terms like the "sunk-cost fallacy" (Arkes and Blumer 1985) and "lock-in" (Cantarelli et al. 2010b) to describe similar phenomena. Escalation of commitment is captured in popular proverbs such as, "Throwing good money after bad" and, "In for a penny, in for a pound."

In its original definition, escalation of commitment is unreflected and non-deliberate. People don't know they are subject to the bias, as with other cognitive biases. However, once you understand the mechanism it may be used deliberately. In his autobiography, Elia Kazan (1997: 412-13), a famous Hollywood director who discovered film star Marlon Brando, explains how he used sunk costs and escalation of commitment to get his projects going:

> "Quickly I planned my position on costs ... My tactic was one familiar to directors who make films off the common path: to get the work rolling, involve actors contractually, build sets, collect props and costumes, expose negative, and so get the studio in deep. Once money in some significant amount had been spent, it would be difficult for Harry [Cohn, President and co-founder of Columbia Pictures] to do anything except scream and holler. If he suspended a film that had been shooting for a few weeks, he'd be in for an irretrievable loss, not only of money but of 'face.' The thing to do was get the film going."

Kazan here combines strategic misrepresentation with cognitive bias to achieve take-off for his projects. The misrepresentation consists in initially (a) being "economical with the truth" regarding the real cost of his projects, and (b) just "get the film going" to sink in sufficient cost to create a point of no return. After this, Kazan trusts the studio head to fall victim to cognitive bias, specifically sunk cost and escalation of commitment, in the grip of which he will allocate more money to the film instead of halting it, which might have been the rational decision. This is the studio head's version of Thaler's "driving into the blizzard," described above. As argued earlier, such interaction between cognitive and political bias is common in shaping project outcomes. Most project managers will know examples similar to Kazan's. It is too simple to think of outcomes as being generated solely by *either* cognitive bias *or* political bias. Such purity may be constructed in lab experiments. In real life, *both* are typically at play with complex interactions between the two.

A number of excellent case studies exist that demonstrate the pertinence of escalation of commitment to project planning and management, e.g., of Expo 86 (Ross and Staw 1986), the Shoreham nuclear power plant (Ross and Staw 1993), and Denver International Airport (Monteagre and Keil 2000), each of which present their own version of "driving into the blizzard."

Above we saw how optimism bias undermines project performance. Escalation of commitment amplifies this. Consider that once a forecast turns out to have been optimistic, often the wisest thing would be to give up the project. But escalation of commitment and the sunk-cost fallacy keep decision makers from doing the right thing. Instead they keep going, throwing good money after bad.

Escalation of commitment often co-exists with and is reinforced by what has been called "preferential attachment" or the "Yule process" (Barabási and Albert 1999, Gabaix 2009, Barabási 2014). Preferential attachment is a procedure in which some quantity, e.g., money or connections in a network, is distributed among a number of individuals or units according to how much they already have, so that those who have much receive more than those who have little, known also as the "Matthew effect."

In project planning and management, Flyvbjerg (2009) argued that the investments that look best on paper get funded, and that these are the investments with the largest cost underestimates and therefore the largest need for additional funding during delivery, resulting in preferential attachment of funds to these investments, once they have their initial funding. After an investment has been approved and funded, typically there is lock-in and a point of no return, after which escalation of commitment follows, with more and more funds allocated to the original investment to close the gap between the original cost underestimate and actual outturn cost (Drummond 2017, Cantarelli et al. 2010b).

Interestingly, preferential attachment has been identified as a causal mechanism that generates outcome distributions with a fat upper tail, specifically power-law distributions (Krapivsky and Krioukov 2008, Barabási 2014). In the case of cost, this would predict an overincidence (compared with the Gaussian) of extreme cost overruns. So far, we have tested the thesis of power-law outcomes for cost and cost overrun with the Olympic Games, where the thesis found strong support in the data (Flyvbjerg et al. 2021). Currently we are further testing the thesis for information technology projects, while tests of other project types are in the pipeline. Should the thesis hold across project types, we may be in the first stages of discovering a general theory of project management, with more fundamental and more scientific explanations of project outcomes than those found in conventional theory.

## Discussion

Scientific revolutions rarely happen without friction. So, too, for the behavioral revolution. It has been met with skepticism, including from parts of the project management community (Flyvbjerg et al. 2018). Some members prefer to stick with conventional explanations of project underperformance in terms of errors of scope, complexity, labor and materials prices, archaeology, geology, bad weather, ramp-up problems, demand fluctuations, etc. (Cantarelli et al. 2010a).

Behavioral scientists would agree with the skeptics that scope changes, complexity, etc. are relevant for understanding what goes on in projects, but would not see them as root causes of outcomes. According to behavioral science, the root cause of, say, cost overrun, is the well-documented fact that project planners and managers keep underestimating scope changes, complexity, etc. in project after project.

From the point of view of behavioral science, the mechanisms of scope changes, complex interfaces, price changes, archaeology, geology, bad weather, business cycles, etc. are not unknown to project planners and managers, just as it is not unknown that such mechanisms may be mitigated. However, project planners and managers often underestimate these mechanisms and mitigation measures, due to optimism bias, overconfidence bias, the planning fallacy, strategic misrepresentation, etc. In behavioral terms, unaccounted for scope changes etc. are manifestations of such underestimation on the part of project planners and managers, and it is in this sense bias and underestimation are *root causes* and scope changes etc. are just *causes*. But because scope changes etc. are more visible than the underlying root causes, they are often mistaken for the cause of outcomes, e.g., cost overrun.

In behavioral terms, the causal chain starts with human bias (political and cognitive) which leads to underestimation of scope during planning which leads to unaccounted for scope changes during delivery which lead to cost overrun. Scope changes are an intermediate stage in this causal chain through which the root causes manifest themselves. Behavioral science tells project planners and managers, "*Your biggest risk is you.*" It is not scope changes, complexity, etc. in themselves that are the main problem; it is how human beings misconceive and underestimate these phenomena, through optimism bias, overconfidence bias, strategic misrepresentation, etc. This is a profound and proven insight that behavioral science brings to project planning and management. You can disregard it, of course. But if you do, project performance would likely suffer. You would be the gambler not knowing the odds of their game.

Behavioral science is not perfect. We saw above how behavioral economics suffers from a "psychology bias," in the sense it tends to reduce behavioral biases to cognitive biases, ignoring political bias in the process, thus committing the very sin it accuses conventional economics of, namely theory-induced blindness resulting in limited rationality. Gigerenzer (2018) goes further and criticizes behavioral economics for "bias bias," and he is right when he calls for conceptual clarification. Not all behavioral biases are well defined, or even well delineated: many and large overlaps exist between different biases that need clarification, including for the ten above. Just as seriously, many biases have only been documented in simplified lab experiments, but are tacitly assumed to hold in real-life situations outside the lab, without sound demonstration that the assumption holds. Finally, the psychology used by behavioral economists is not considered cutting-edge by psychologists, a fact openly acknowledged by Thaler (2015: 180), who further admits it is often difficult to pin down which specific behavioral bias is causing outcomes in a given situation or to rule out alternative explanations (ibid. 295).

Nevertheless, the behavioral revolution seems to be here to stay and it entails an important change of perspective for project management: The problem with project cost overruns and benefit shortfalls is not error but bias, and as long as we try to solve the problem as something it is not (error), we will not succeed. Estimates and decisions need to be de-biased, which is fundamentally different from eliminating error. Furthermore, the problem is not even cost overruns or benefit shortfalls, it is cost underestimation and benefit overestimation. Overrun, for instance, is mainly a consequence of underestimation, with the latter happening upstream from overrun, for big projects often years before overruns manifest. Again, if we try to solve the problem as something it is not (cost overrun), we will fail. We need to solve the problem of upstream cost underestimation in order to solve the problem of downstream cost overrun. Once we understand these straightforward insights, we understand that we and our projects are better off with an understanding of behavioral science and behavioral bias than without it.

**Endnotes**

[1] Specifically, Kahneman met with and interviewed the Chair for Transportation at RPA, Jeff Zupan, and his team.

---

[2] Table 2 shows the results from the largest dataset of its kind. Other studies, based on smaller datasets, show similar results, e.g., Albalate and Bel (2014), Altshuler and Luberoff (2003), Bain (2009), Bain and Polakovic (2005) Bain and Wilkins (2002), Caro (1975), Delaney and Eckstein (2003), Feynman (2006: 397-407; 2007a,b), Flyvbjerg (2005), Flyvbjerg et al. (2002), Goldblatt (2016), Kain (1990), Pickrell (1992), Richmond (2005), Winch (2008), and Zimbalist (2015).